\documentclass[11pt,twoside]{article}
\usepackage{jltp_latex2e}

\title{Vortex Stability in a Trapped Bose Condensate}

\author{Alexander L.~Fetter\address{Departments of Physics and Applied
Physics, Stanford University, Stanford, CA 94305-4060, USA}}

\runninghead{A.~L.~Fetter}{Vortex Stability}

\begin{document}

\begin{abstract}
A vortex in a trapped Bose-Einstein condensate can experience at least two
 types of instabilities. (1).  Macroscopic
hydrodynamic motion of the vortex core relative to the center of mass of
the condensate requires some  process to dissipate energy. (2).  Microscopic
small-amplitude normal modes can also induce an instability.   In one specific
example, the vortex core again moves relative
to the overall center of mass, suggesting that there may be only   a single
  physical mechanism.

\end{abstract}

\maketitle

\section{INTRODUCTION}
The  remarkable recent experimental creation  of
 Bose-Einstein condensates  in trapped low-temperature
alkali gases~\cite{And,Davis,Brad} has generated great interest   in the
possibility of vortex states.  Such vortices have been widely
studied in superfluid ${}^4$He,\cite{Hess,PS} but no clear experimental
evidence demonstrates the existence of  a vortex in a trapped Bose
condensate.   Theoretical work  has
concentrated on the critical angular velocity $\Omega_{c1}$ for vortex
creation~\cite{Dal},  the normal modes of a condensate containing a
vortex~\cite{Sinha,Dodd,SF,ZS}, and general considerations of
stability.~\cite{Rokhsar}

	The present work studies two specific types of instabilities of a
vortex in
a trapped Bose condensate.  The first (Sec.~2) is a hydrodynamic instability
involving the macroscopic motion of a vortex relative to the background
condensate, including the effect of the nonuniform condensate density, which
has not previously been incorporated.~\cite{Hess,PS,Rokhsar} The second
instability (Sec.~3) arises from the microscopic internal oscillations  of the
vortex, which has been studied in a particular geometry by Dodd et
al.~\cite{Dodd,Rokhsar}.  This latter behavior is especially clear in the
weak-coupling limit, where the normal modes of the vortex
differ qualitatively  from those of a vortex-free condensate.

\section{HYDRODYNAMIC INSTABILITY  OF A VORTEX}

The behavior of a trapped condensate depends crucially on the number $N$ of
atoms in the condensate. Consider an axisymmetric trap with a potential
\begin{equation}
V_{\rm tr}= \frac{1}{2}M(\omega_\perp^2\rho^2 + \omega_z^2z^2),
\end{equation}
where $M$ is the atomic mass and  ($\rho,\phi,z$) are the familiar
cylindrical polar coordinates.  The radial and axial oscillator lengths
$d_\perp =
\sqrt{\hbar/M\omega_\perp}$ and $d_z = \sqrt{\hbar/M\omega_z}$  characterize
the condensate's dimensions for an ideal trapped Bose gas (the corresponding
volume is of order   $d_0^3 \equiv d_\perp^2d_z$).   The
short-range two-body interaction potential may be written as
\begin{equation}
V({\bf r}) \approx g\delta^{(3)}({\bf r}),
\end{equation}
where $g \approx 4\pi a \hbar^2/M$ relates the coupling strength to
 the $s$-wave scattering length $a$ (here assumed positive).  The
nonuniform condensate wave function $\psi({\bf r})$ obeys the Gross-Pitaevskii
(GP)  equation~\cite{EPG,LPP}
\begin{equation}
\left(T+ V_{\rm tr} + V_H\right)\psi = \mu\psi\label{GP},
\end{equation}
where $T=-\hbar^2\nabla^2/2M$ is the kinetic-energy operator, $V_H({\bf
r}) = gN|\psi({\bf r})|^2=4\pi aN \hbar^2|\psi({\bf r})|^2/M$  is the Hartree
potential energy of one particle with the remaining particles, and $\mu$ is
the chemical potential.

Typically, the dimensionless ratio $a/d_0$ is small (of order $10^{-3}$),
but   the relevant dimensionless parameter~\cite{BP}
$ Na/d_0$ varies linearly with  the condensate number $N$. If $Na/d_0$  is
small, then the interactions are weak,  and the condensate acts
like a nearly ideal Bose gas with characteristic radial and axial dimensions
$d_\perp$ and $d_z$.  In the opposite limit $Na/d_0
\gg 1$, the repulsive interactions predominate, and the condensate expands
well beyond the harmonic-oscillator lengths.  In this  Thomas-Fermi (TF)
limit, the kinetic energy is negligible, and the condensate density follows
from the GP equation
\begin{equation}
N|\psi|^2 = g^{-1}\left(\mu - V_{\rm tr}\right) =
n(0)\left(1-\frac{\rho^2}{R_\perp^2} - \frac{z^2}{R_z^2}\right),
\end{equation}
where $n(0)= \mu/g$ is the central density and $R_\perp$ and $R_z$ are the
 radial and axial condensate dimensions with  $R_\alpha^2/d_\alpha^2  =
2\mu/\hbar\omega_\alpha\gg1$ for $\alpha = \perp$ and $z$.  In the TF
limit, the normalization condition $\int dV\,|\psi|^2 = 1$ yields the
dimensionless parameter
\begin{equation}
\frac{Na}{d_0} = \frac{1}{15}\frac{R_0^5}{d_0^5}\gg 1,
\end{equation}
where $R_0^3 = R_\perp^2R_z$ characterizes the  TF condensate volume.  The
repulsive interactions also introduce yet another important length
[the coherence length
$\xi
\equiv 1/\sqrt{8\pi n(0)a}\,$], which   here determines the vortex-core
radius;  in the TF limit, it obeys the relation
$\xi R_0 = d_0^2$, implying the set of inequalities $\xi\ll d_0\ll R_0$.

If the trap rotates with angular velocity $\Omega$, the relevant ``free
energy'' is $F = E-\Omega L_z$, where $E$ is the
 energy of the condensate and $L_z$ is its angular momentum.   The
phase $S$  of the condensate wave function determines the velocity ${\bf v} =
(\hbar/M)\nabla S$, and the GP equation (\ref{GP}) provides an explicit
expression for the TF free energy
\begin{equation}
F\approx \int dV\,\left({\textstyle{\frac{1}{2}}}Mn\,v^2 + nV_{\rm tr} +
{\textstyle{\frac{1}{2}}}gn^2 -Mn\Omega\hat z\cdot {\bf r\times
v}\right),\label{free}
\end{equation}
where  spatial derivatives of the density are neglected.

This integral for $F$ provides a variational expression in the rotating
frame, and the classical velocity potential $\Phi$ provides a convenient
approximation for the phase $S$.  The simplest case is a   uniform fluid with
density $\overline n$ per unit length in a long circular cylinder with radius
$R$.  When a vortex with circulation $\kappa = h/M$ is added to the system at
a distance
$x_0 R<R $ from the  symmetry axis, the change in the free energy
$\Delta \overline F$ is~\cite{Hess,PS}
\begin{equation}
\Delta \overline F = \frac{M\kappa^2\overline
n}{4\pi}\left[\ln\left(\frac{R}{\xi}\right)+\ln\left(1-x_0^2\right)
-\frac{\Omega}{\Omega_0}\left(1-x_0^2\right)\right],
\end{equation}
where $\Omega_0 = \kappa/2\pi R^2 = \hbar/MR^2$ is a characteristic angular
speed.   In the absence of dissipation, the vortex executes a circular
orbit at fixed radius under  the influence of its opposite image at a
distance
$R/x_0>R$ from the axis.  In the presence of dissipation, however, the vortex
moves to reduce its free energy.  For
$\Omega <\Omega_0$, the free energy
$\Delta \overline F$ decreases monotonically with increasing $x_0$, so the
vortex simply spirals outward and annihilates with its image.  In contrast,
if $\Omega>\Omega_0$, the free energy near the axis increases with increasing
$x_0$, so that a vortex near the axis tends to return to  the
center of the container. This situation is merely metastable for $\Omega\le
\Omega_{c1}
\equiv
\Omega_0\ln(R/\xi)$, since the free energy at the center is higher than that
at the wall (with a barrier at some intermediate distance), but the
vortex becomes a true equilibrium state when $\Omega\ge \Omega_{c1}$.

The TF  radial profile density $n = n(0)(1-\rho^2/R^2)$ provides a  more
realistic description of a nonuniform rotating fluid in a long circular
container.  The mean density per unit length $\overline n$ is half the
central density $n(0)$, and the corresponding  vortex-induced change in the
free energy  is
\begin{equation}
\Delta F = \frac{M\kappa^2\overline
n}{4\pi}\,(1-x_0^2)\left[2\ln\left(\frac{R}{\xi}\right)
+\frac{1+x_0^2}{x_0^2}\ln\left(1-x_0^2\right)
-\frac{\Omega}{\Omega_0}\left(1-x_0^2\right)\right].
\end{equation}
An expansion for small $x_0$ shows that a central vortex is
stable for $\Omega>\Omega_{c1} \equiv \Omega_0[2\ln(R/\xi)-1]$,  unstable
for $\Omega<\Omega_{m}\equiv \Omega_0[\ln(R/\xi) + \frac{1}{4}] =
\frac{1}{2}\Omega_{c1}+\frac{3}{4}$, and metastable for
$\Omega_{m}<\Omega<\Omega_{c1}$.  Thus, inclusion of a realistic density
profile affects the detailed form of the free energy (the metastable region is
considerably reduced), but the qualitative picture remains unchanged.

\section{STABLE AND UNSTABLE MODES OF A  VORTEX}

Rokhsar~\cite{Rokhsar} has argued that a vortex in a nearly ideal trapped
condensate is unstable because the relatively large core (of order
$d_\perp$) implies a bound state in the core.  In particular, he noted that
Dodd et al.~\cite{Dodd} found an unstable solution of the Bogoliubov
equations for a condensate containing a singly quantized vortex (this
unstable mode is distinct from the rigid dipole modes that
necessarily oscillate at  the trap frequency $\omega_\perp$ independent  of
the interaction parameter
$Na/d_0$).

To clarify the physical meaning of this instability, it is helpful to study
the relevant solutions of the Bogoliubov equations
\begin{eqnarray}
(T+V_{\rm tr}-\mu + 2gN|\psi|^2)u_j -gN(\psi)^2v_j = \hbar
\omega_ju_j;\label{Boga}\\ -gN(\psi^*)^2u_j+(T+V_{\rm tr}-\mu +
2gN|\psi|^2)v_j  = -\hbar\omega_jv_j,\label{Bogb}
\end{eqnarray}
where $u_j$ and $v_j$ are the normal-mode amplitudes and $\omega_j$ is the
corresponding frequency.  A state $j$ with positive norm [$\int
dV(|u_j|^2-|v_j|^2) = 1$] is potentially unstable if $\omega_j<0$ since the
creation of a quasiparticle in the $j$th mode lowers  the energy  relative to
that of the condensate~\cite{ALF}.  For any solution of the Bogoliubov
equations, the perturbations in the particle density and velocity potential
are

\begin{equation}
n_j' = \left(\psi^*u_j-\psi v_j\right)\,e^{-i\omega_jt},\qquad \Phi'_j =
\frac{\hbar}{2Mi|\psi|^2}\left(\psi^* u_j + \psi
v_j\right)\,e^{-i\omega_jt}.\label{pert}
\end{equation}

For a nearly ideal gas, the terms proportional to $gN$ in Eqs.~(\ref{GP}),
(\ref{Boga}), and (\ref{Bogb}) can be treated in perturbation theory. If the
condensate has no vortex, then the leading approximation to the condensate
wave function is
$\psi\approx \chi_{00}$, where  I   ignore the $z$-dependent part of the
wave function and $\chi_{n_+,n_-}(\rho,\phi)$ is  a normalized
two-dimensional  oscillator wave function containing $n_+$ and
$n_-$ right and left circular quanta created by the raising
operators $a_\pm^\dagger\equiv a_x^\dagger \pm ia_y^\dagger$.~\cite{CT} The
lowest excited solutions of the Bogoliubov equations for this ground-state
vortex-free condensate are the rigid dipole modes with frequency
$\omega_\pm =
\omega_\perp$ for all interaction strengths.  Apart from corrections of order
$g$, their detailed form
\begin{equation}
u_+ \approx \chi_{10},\quad u_-\approx
\chi_{01},\quad v_\pm\approx 0,\quad
\hbox{with $\omega_\pm = \omega_\perp$}
\end{equation}
follows either by direct
construction or from the general explicit dipole solutions of the Bogoliubov
equations~\cite{FR}

\begin{equation}
u_\pm = a_\pm^\dagger \psi,\quad v_\pm = a_\mp \psi^*,\label{dipole}
\end{equation}
where $\psi$ is {\it any\/} solution of the GP equation (\ref{GP}).
  The corresponding density and velocity-potential
perturbations follow from Eqs.~(\ref{pert})

\begin{equation}
 n_\pm' \approx n_0\,\rho\,e^{\pm i\phi}e^{-i\omega_\perp t};\quad \Phi_\pm'
\approx\frac{1}{2i}\,\rho \,e^{\pm i\phi}e^{-i\omega_\perp t},
\end{equation}
where $n_0=|\chi_{00}|^2$ is the unperturbed condensate density.
They represent a circular motion of the rigidly  displaced condensate in the
positive and negative sense, respectively.

The situation is much more interesting for a singly quantized vortex,  with
condensate wave function $\psi \approx
\chi_{10}\propto e^{i\phi}\,\rho\, e^{-\rho^2/2}$ in the
noninteracting limit.  The general construction in Eq.~(\ref{dipole}) now
yields

\begin{equation}
\pmatrix{u_+\cr v_+} = \pmatrix{\sqrt 2\,\chi_{20}\cr
\chi_{00}},\quad\pmatrix{u_-\cr v_-} = \pmatrix{\chi_{11}\cr
0}, \quad\hbox{with $\omega_\pm = \omega_\perp$}
\end{equation}
so that the $+$ state involves a coherent superposition of a particle and a
hole, differing from that for the $-$ state.  Nevertheless, they both
oscillate  with frequency
$\omega_\perp$, and the density  and
velocity-potential perturbations [from Eqs.~(\ref{pert})] have the same
 form~\cite{SF}
\begin{equation}
n_\pm' \approx n_v\left(\rho -
\frac{1}{\rho}\right)e^{\pm i \phi}\,e^{-i\omega_\perp t},\quad \Phi_\pm'
\approx \frac{\hbar}{2Mi}\,\left( \rho \pm
\frac{1}{\rho}\right)e^{\pm i \phi}\,e^{-i\omega_\perp t}
\end{equation}
where $n_v \approx |\chi_{10}|^2$ is the condensate density for
the vortex state (this expression also follows by an expansion of the
condensate  density for a small rigid displacement of both the center of mass
and the position of the  vortex core).

The present case of a singly quantized vortex leads to an additional
anomalous mode with  frequency
$\omega_a
\approx -\omega_\perp$, reflecting the single-particle transition from
the vortex state to the (lower)  Gaussian ground state
$\chi_{00}$.~\cite{Rokhsar}  To zero order in the interaction parameter
$aN/d_z$, the associated Bogoliubov amplitudes are
$u_a\approx
\cosh\theta\,\chi_{00}$ and
$v_a\approx
\sinh\theta\,\chi_{02}$, which are properly normalized for any value of the
parameter $\theta$.  To determine the actual value of $\theta$, it is
necessary to use first-order perturbation theory,  yielding the expressions
\begin{equation}\pmatrix{u_a\cr v_a} = \pmatrix{\sqrt 2\,\chi_{00}\cr
\chi_{02}}\quad\hbox{and} \quad
\omega_a\approx \omega_\perp \left(-1+
\frac{aN}{4d_z}\sqrt{\frac{2}{\pi}}\right).
\end{equation}
Like the $+$ dipole mode of the vortex, this anomalous  mode is  a
coherent superposition of a particle and a hole.  Note that the
frequency here {\it differs\/} from the trap frequency  (and hence from the
rigid dipole-oscillation frequency) for
any nonzero interaction strength.~\cite{Dodd} The corresponding density and
velocity-potential perturbation are
\begin{equation}
n_a' \approx -\frac{n_v}{\sqrt 2}\left(\rho -
\frac{2}{\rho}\right)e^{-i\phi}e^{-i\omega_a
t}\quad\hbox{and}\quad \Phi_a' \approx
\frac{\hbar}{2\sqrt 2
Mi}\left(\rho+\frac{2}{\rho}\right)e^{-i\phi}e^{-i\omega_a
t}.\label{dens}\end{equation}
 A detailed study
for this anomalous mode shows that the position of the vortex core is shifted
by twice that of the condensate's center of mass.  In addition, the overlap
integral
$\int d^2\rho\,n_a' (x\pm i y )
$ vanishes identically, so this mode is not excited by the dipole oscillation
of the center of mass.

It is striking that both types of instability (hydrodynamic in Sec.~2
and microscopic in Sec.~3) involve the relative motion of the vortex
core and the center of mass;~\cite{Rokhsar}  they may well be  two
descriptions of the same physics.   It will be  particularly interesting to
study the character and role  of this anomalous
  mode  for increasing coupling strength.

\section*{ACKNOWLEDGMENTS}
I thank M.~Edwards, D.~Rokhsar, G.~Shlyapnikov, and S.~Stringari for
many  discussions.  This research is supported by NSF grant DMR
94-21888.

\end{document}